\documentclass[A4paper,prl,twocolumn,amsmath,amssymb,showpacs,superscriptaddress,floatfix]{revtex4-1}

\usepackage{graphicx}\usepackage[hypertex]{hyperref}
\usepackage{srcltx} %
\usepackage{srctex}
  \DeclareMathOperator{\Imag}{Im}\DeclareMathOperator{\ch}{ch}
\DeclareMathOperator{\diag}{diag}

\begin{document}
\newcommand{\remark}[1] {\noindent\framebox{
\begin{minipage}{0.96\columnwidth}\textbf{\textit{ #1}}
\end{minipage}}
}
\newcommand{\bnabla}{\boldsymbol{\nabla}} \newcommand{\Tr}{\mathrm{Tr}} \newcommand{\Dk}{\check{\Delta}_{\cal K}} \newcommand{\Qk}{\check{Q}_{\cal K}} \newcommand{\Fk}{\check{\Phi}_{\cal K}} \newcommand{\Ak}{\check{\mathbf{A}}_{\cal K}} \newcommand{\Si}{\check{\Xi}} \newcommand{\cK}{\cal K} \newcommand{\bz}{{\mathbf z}} \newcommand{\bx}{{\mathbf x}} \newcommand{\br}{{\mathbf r}}\newcommand{\bG}{{\mathbf G}} \newcommand{\bu}{{\mathbf u}} \newcommand{\bq}{{\mathbf q}} \newcommand{\cH}{{\cal H}} \newcommand{\dif}{{\mathrm d}}

%-------------- Principle value
\def\Xint#1{\mathchoice
{\XXint\displaystyle\textstyle{#1}}%
{\XXint\textstyle\scriptstyle{#1}}%
{\XXint\scriptstyle\scriptscriptstyle{#1}}%
{\XXint\scriptscriptstyle\scriptscriptstyle{#1}}%
\!\int}
\def\XXint#1#2#3{{\setbox0=\hbox{$#1{#2#3}{\int}$}
\vcenter{\hbox{$#2#3$}}\kern-.55\wd0}}
\def\ddashint{\Xint=}
\def\dashint{\Xint-}
%--------------

\title{Out-of-equilibrium heating of electron liquid: fermionic and bosonic temperatures}

\author{A.~Petkovi\'{c}}
\affiliation{Materials Science Division, Argonne National Laboratory, Argonne, Illinois 60439, USA}
\author{N.\,M.~Chtchelkatchev}
\affiliation{Materials Science Division, Argonne National Laboratory, Argonne, Illinois 60439, USA}
\affiliation{Institute for High Pressure Physics, Russian Academy of Sciences, Troitsk 142190, Moscow region, Russia}
\affiliation{L.D. Landau Institute for Theoretical Physics, Russian Academy of Sciences,
Moscow 117940, Russia}
\author{T.\,I.~Baturina}
\affiliation{Materials Science Division, Argonne National Laboratory, Argonne, Illinois 60439, USA}
\affiliation{Institute of Semiconductor Physics, 13 Lavrentjev Avenue, Novosibirsk 630090, Russia}
\affiliation{Novosibirsk State University, 2 Pirogova Street, Novosibirsk 630090, Russia}
\author{V.\,M.~Vinokur}
\affiliation{Materials Science Division, Argonne National Laboratory, Argonne, Illinois 60439, USA}

\date{\today}

\begin{abstract}
We investigate out-of-the equilibrium properties of the electron liquid in a
two-dimensional disordered superconductor subject to the electric bias and
temperature gradient. We calculate kinetic coefficients and Nyquist noise, and find that
they are characterized by distinct effective temperatures: $T_{\mathrm{e}}$, characterizing
single-particle excitations, $T_{\mathrm{Cp}}$,  describing the Cooper pairs,
and $T_{\mathrm{eh}}$, corresponding to electron-hole or dipole excitations.
Varying the ratio between the electric $j$ and thermal $j_{\mathrm{th}}$ currents and boundary conditions one
can heat different kind of excitations tuning their corresponding temperatures.
We propose the experiment to determine these effective temperatures.
\end{abstract}


%73.23.-b  Electronic transport in  mesoscopic systems
%74.45.+c Proximity effects; Andreev effect; SN and SNS junctions
%74.50.+r Tunneling phenomena; point contacts, weak links, Josephson effects
%74.78.Fk Multilayers, superlattices, heterostructures

\maketitle

A temperature, the term  quantifying the common idea of ``hot" and ``cold,"
is one of the most fundamental concepts in physics.
In statistical physics, the temperature is a parameter that controls the probability of
the energy distribution of a given system over the possible states.
The basic property of the temperature is that in an equilibrium it is the constant
all over the system involved due to the interactions between the constituent subsystems.
In out of the equilibrium case, different subsystems can acquire different temperatures.
Electrons and phonons which equilibrate within themselves,
but can possess different temperatures if weakly coupled, is an exemplary case of such
a non-equilibrium situation~\cite{Giazotto,Volkov_Kogan}.

In this Letter we show that strong correlations bring more richness to the
non-equilibrium physics.  We construct a kinetic theory of
electronic transport in a disordered two-dimensional (2D) superconductor
and demonstrate that distinct electronic subsystems
respond differently to external drive.
We derive transport kinetic coefficients and find that
they are controlled by distinctive effective fermionic
and bosonic temperatures each corresponding to
a particular electronic subsystems: $T_{\mathrm{e}}$ for the single-particle
excitations, $T_{\mathrm{\scriptscriptstyle Cp}}$ for Cooper pairs, and $T_{\mathrm{eh}}$
for electron-hole
pairs (dipoles).  These temperatures are defined through the respective
energy distribution functions, and the relations between them are determined by the
character of the external drive, or more specifically, by the ratio of its components,
the applied voltage bias and/or temperature gradient (see Fig.\,1).

%----------------------
\begin{figure}[h]
\includegraphics[width=0.95\columnwidth]{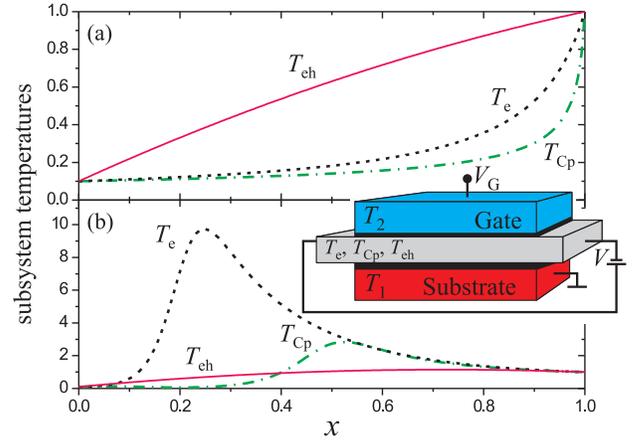}\\
\caption{Temperatures of electron-hole pairs
($T_{\mathrm{eh}}$), Cooper pairs ($T_{\mathrm{\scriptscriptstyle Cp}}$)
and quasi-particles ($T_{\mathrm{e}}$) vs. the parameter $x=R_2/(R_1+R_2)$,
characterizing the ratio between tunneling resistances $R_1$
between the substrate and superconducting film and $R_2$, between the film and the gate.
The substrate temperature is  $T_1$ and the gate temperature is $T_2$,
$T_2/T_1=0.1$, the voltage drop between them is $V_{\scriptscriptstyle G}$.
(a) $V_{\scriptscriptstyle G}=0$, where only thermal current is present;
(b) $eV_{\scriptscriptstyle G}/T_1=3$, with $V_G j\gtrsim j_{\mathrm{th}}$. \\ Generally,
in the film driven far from equilibrium and for $R_1\simeq R_2$,
different interrelations between fermionic and bosonic temperatures can be realized:
$T_{\mathrm{e}}\simeq T_{\mathrm{\scriptscriptstyle Cp}}\ll T_{\mathrm{eh}}$
for $j_{\mathrm{th}}\gg V_{\scriptscriptstyle G} j$,
while $T_{\mathrm{e}}\simeq T_{\mathrm{Cp}}\gg T_{\mathrm{eh}}$
for $j_{\mathrm{th}}\ll V_{\scriptscriptstyle G} j$.
\label{fig1}}
\end{figure}
%----------------------

We consider 2D disordered systems such that the Thouless energy corresponding to
diffusion across the film  $E_T^\perp=D/d^2$,
where $D$ is the diffusion coefficient, $d$ is the film thickness,
well exceeds all the fermionic and bosonic temperatures.
Hereafter we will use units where $k_B=\hbar=c=1$ ($c$ is the light velocity).
The bias-dependent subsystem temperatures manifest themselves most
profoundly in the leading quantum corrections~\cite{Varlamov_book} to
the sheet conductance
in the singlet, $\delta G^{\mathrm {\scriptscriptstyle AA}}$ (Aronov-Altshuler correction)
and in the Copper channel, $\delta G^{\mathrm{\scriptscriptstyle MT}}$
(Maki-Thomson correction) close to the transition:
\begin{align}\label{eq:1}
\delta G^{\mathrm{\scriptscriptstyle AA}}&=\frac{e^2}{2\pi^2  }\ln({T_{\mathrm{eh}}\tau})\,,
\\\label{eq:2}
\delta G^{\mathrm{\scriptscriptstyle MT}}&=\frac{e^2}{\pi  }{T_{\mathrm{\scriptscriptstyle Cp}}\tau_{\mathrm{\scriptscriptstyle GL}}}
\ln{\left(\tau_{\phi}/\tau_{\mathrm{\scriptscriptstyle GL}}\right)},
\end{align}
and in the leading contribution to the in-plane Nyquist noise,
\begin{gather}\label{eq.3}
S_{\mathrm {th}}(\omega=0)\approx 2\sigma_{\mathrm{\scriptscriptstyle D}} T_{\mathrm{eh}}.
\end{gather}
Here $e$ is the electron charge, $\tau$ is the elastic scattering time,
$\tau_\phi$ is the dephasing time,
$\tau_{\mathrm {\scriptscriptstyle GL}}$ is the  Ginzburg-Landau time
indicating how close the system is to the superconducting transition, and
$\sigma_{\mathrm{\scriptscriptstyle D}}$ is the Drude conductivity.
We consider disordered system where $T_{\mathrm {eh({\scriptscriptstyle Cp})}}\ll\tau^{-1}$,
$e^2/(\sigma_{\mathrm {\scriptscriptstyle D}}d)\ll 1$, and
$\tau_{\mathrm {\scriptscriptstyle GL}}\ll \tau_{\phi}$.

The fermionic and bosonic  temperatures
are defined through the respective distribution functions,
that are assumed to be space and time independent.
We start with the fermionic temperature $T_{\mathrm e}$ describing
the low lying single-particle excitations.
In an equilibrium, the electron (hole) gauge invariant
distribution function in the disordered electron liquid is
$\tilde f_{\mathrm{e(h)}}^{\mathrm {(eq)}}(\epsilon)
=f_{\mathrm{\scriptscriptstyle F}}(\epsilon)\equiv 1/[\exp(\epsilon/T)+1]$,
where $\epsilon$ is the quasiparticle energy,
and the temperature appears as its low-energy ``cut-off'':
$T=\epsilon/2[1-2\tilde f_{\mathrm {e(h)}}^{\mathrm {(eq)}}(\epsilon)]|_{\epsilon\to 0}$.
Generalizing  onto an  out of the equilibrium case, one arrives at:
\begin{gather}\label{eq:Te_def}
    T_{\mathrm e}=\{-4\lim_{\epsilon\to0}\partial_\epsilon
\tilde f_{\mathrm {\scriptscriptstyle L}}(\epsilon)\}^{-1},
\end{gather}
where $\tilde f_{\mathrm {\scriptscriptstyle L}}(\epsilon)
=[\tilde f_{\mathrm e}(\epsilon)+\tilde f_{\mathrm h}(\epsilon)]/2$, and
$\tilde f_\mathrm{e(h)}(\epsilon)$
is a general nonequilibrium electron (hole) distribution function.

The electron-hole ($e$-$h$) pair temperature $T_\mathrm{eh}$ emerges in the bosonic distribution
function of the electron-hole excitations mediating, for example, the
electronic transport in metals~\cite{CVB}.
The bosonic nature of the $e$-$h$
environment is revealed by the fluctuation-dissipation theorem (FDT)
for the density-density correlation function in a quasi-equilibrium regime:
 $\Pi^K(\mathbf{q},\omega)=2i (1+2N^{\rm(eq)}_\omega) \Imag\Pi^R(\omega,\mathbf{q})$,
 where $N^{\rm(eq)}_\omega$ is the Bose distribution function of the
 $e$-$h$ excitations, $\Pi^R(x-y)$ is the retarded correlator of the electron density, and
 $\Pi^K(x-y)$ is the Keldysh density correlator.
 The low-energy cut-off of the
 distribution function $N^{\rm(eq)}$ defines $T_{\mathrm{eh}}=\omega N^{\rm(eq)}_\omega|_{\omega\to 0}$.
In
 the nonequilibrium case we find
\begin{gather}\label{eq:Teh_def}
    T_{\mathrm{eh}}=\lim_{\omega\to 0}\omega N_\omega,
    \\ \label{eq:1223}
    N_\omega=\frac1{2\omega}\sum_{\alpha=e,h}\int
[1-\tilde f_\alpha(\epsilon_-)]\tilde f_\alpha(\epsilon_+) d\epsilon,
\end{gather}
where $\epsilon_\pm=\epsilon\pm\omega/2$.
The integrand in \eqref{eq:1223} gives the probability for the state with the
energy $\epsilon_+$ be occupied and the state with the energy $\epsilon_-$ be empty.
In an equilibrium $N_\omega\to N^{\rm(eq)}_\omega$.

Turning to the nonequilibrium generalization of the FDT in the Cooper channel, i.e.
taking the Keldysh component of the fluctuation propagators in the Cooper channel,
one arrives at the temperature for the fluctuating Cooper pairs as
\begin{gather}\label{eq:Tcp}
    T_{\mathrm{\scriptscriptstyle \scriptscriptstyle Cp}}=2T_{\mathrm e} z_{\mathrm{\scriptscriptstyle Cp}}\left\{\tilde f_{\mathrm e}(0)
[1-\tilde f_{\mathrm h}(0)]+\tilde f_{\mathrm h}(0)[1-\tilde f_{\mathrm e}(0)]\right\}.
\end{gather}
Here  $z_{\mathrm{Cp}}= (1+(T_{\mathrm{e}}/\Omega)^2)^{-1}$ and
$\Omega^{-1}=4 \dashint_\epsilon [\tilde f_{\mathrm{\scriptscriptstyle R}}
(0)-\tilde f_{\mathrm{\scriptscriptstyle R}}
(\epsilon)]/\pi\epsilon^2$,
where $\tilde f_{\mathrm{\scriptscriptstyle R}}(\epsilon)
=\tilde f_{\mathrm e}(-\epsilon)-\tilde f_{\mathrm h}(\epsilon)$.
In an equilibrium $T_{\mathrm e}=T$,
$z_{\mathrm{\scriptscriptstyle Cp}}=1$,
$f_{\scriptscriptstyle{\mathrm F}}(0)[1-f_{\mathrm{\scriptscriptstyle F}}(0)]=1/4$,
and therefore $T_{\mathrm{\scriptscriptstyle Cp}}=T$.
The combination of distribution functions appearing in the definition of
$T_{\mathrm{\scriptscriptstyle Cp}}$ can be viewed
as the probability for generating a Cooper pair.

%%%%%%%%%%%
%%%%%%%%%%%
We consider a setup shown in Fig.~\ref{fig1}, where the superconducting film
is sandwiched between the substrate and the gate and is separated from them
by the tunnel barriers with the resistances $R_1$ and $R_2$, respectively.
We can manipulate the density matrix of the system  applying the gate bias,
$V_{\scriptscriptstyle G}$, and changing the temperatures $T_1$ and $T_2$
of the substrate and the gate.
The electronic distribution functions defining the subsystem temperatures in the film
are found from the respective kinetic equations.
The thickness of the superconducting film $d$ is much smaller
than the inelastic quasiparticle scattering length, thus the corresponding
collision integral can be omitted.
Then the kinetic equations in the stationary situation assume
the form of the current conservation law,
 ${\bnabla} \mathbf{j}_\mathrm{e(h)}(\epsilon)=0$,
where the spectral quasiparticle current $\mathbf{j}_\mathrm{e(h)}(\epsilon)=-D{\bnabla}  f_\mathrm{e(h)}$.
Here we use the gauge where $\mathbf{A}=0$.
The boundary conditions require the conservation
of $\mathbf{j}_\mathrm{e(h)}(\epsilon)$ and imply that
$\mathbf{j}_\mathrm{e(h)}(\epsilon)=[f_\mathrm{e(h)}^{(f)}(\epsilon)- f_\mathrm{e(h)}^{(s)}(\epsilon)]/\nu e^2 R_1$,
 where the indexes $s$ and $f$ denote substrate and film, respectively,
and $\nu$ is the single-particle density of states at the Fermi level per spin
projection~\cite{Kuprianov+}.
Similar relation holds for $\mathbf{j}_\mathrm{e(h)}(\epsilon)$
between the film and the gate.
%%%%%%%%%%%%
%%%%%%%%%%%%
The gauge invariant distribution functions satisfy the kinetic equations which
contain directly the electromagnetic fields rather then their respective potentials.
In our case electric field enters the kinetic equation through the boundary conditions.
In order to obtain the gauge-invariant distribution functions,
one uses the transformation
$\tilde f_{\mathrm{e(h)}}(\epsilon)=f_{\mathrm{e(h)}}
(\epsilon\pm e\phi_{{\mathrm f}})$,
where $\phi_{{\mathrm f}}$ is the electric potential in the film, see,
e.g., \cite{Rammer-Smith,Kopnin_book}.
Thus,
\begin{multline}
\tilde f_\mathrm{e(h)}(\epsilon)=xf_{\mathrm{\scriptscriptstyle F}}
\left({\epsilon\pm e V_{\scriptscriptstyle G}(1-x)},{T_1}\right)+
\\
(1-x)f_{\mathrm{\scriptscriptstyle F}}\left({\epsilon\mp e V_{\scriptscriptstyle G} x},{T_2}\right),
\end{multline}
where the temperature appears as the second argument in the Fermi-functions,
and $x=R_2/R_{\scriptsize {\scriptscriptstyle\Sigma}}$
with $R_{\scriptscriptstyle \Sigma}=R_1+R_2$.
Expressing subsystem temperatures through the contact parameters
and using definitions Eqs.\eqref{eq:Te_def}-\eqref{eq:Teh_def}, one finds:
\begin{gather}\label{eq:Teh}
\begin{split}
T_{\mathrm{eh}}\approx x^2T_1 +(&1-x)^2 T_2+
    \\
& x(1-x)eV_{\scriptscriptstyle G} \coth\frac {e V_{\scriptscriptstyle G}}{\sqrt{2(T_1^2+T_2^2)}},
\end{split}
\\\label{eq:Te}
T_\mathrm{e}=\left(\frac {x}{T_1\ch^2\frac {(1-x)e V_{\scriptscriptstyle G}}{2T_1}}
+\frac {(1-x)}{T_2\ch^2\frac {x e V_{\scriptscriptstyle G}}{2T_2}}\right)^{-1}.
\end{gather}
The temperature $T_\mathrm{\scriptscriptstyle Cp}$ is given by Eq.\,\eqref{eq:Tcp},
where $\Omega^{-1}$ becomes $2\mathrm{Im}\{T_2 x \Psi '[1/2-ieV_{\scriptscriptstyle G} (1-x)/(2\pi T_1)]
+T_1 (1-x) \Psi'[1/2+ieV_{\scriptscriptstyle G} x/(2\pi T_2)]\}/(T_1 T_2 \pi^2)$,
and $\Psi$ is the digamma function.
The Ginzburg-Landau time is defined as
$\tau_{\mathrm{\scriptscriptstyle GL}}^{-1} =8z_{\mathrm{Cp}}T_{\mathrm e}\dashint_{\epsilon}[\tilde f_{\scriptscriptstyle{\mathrm R}}
(\epsilon)-\tilde{f}_{\scriptscriptstyle{\mathrm R}}(0)-f_{\scriptscriptstyle{\mathrm F}}(\epsilon,T_{\mathrm c})+1/2]/{\epsilon\pi} +4z_{Cp}T_e^2(1-2\tilde{f}_{\scriptscriptstyle{\mathrm R}}(0))/\Omega$.
At zero voltage, $\tau_{\mathrm{\scriptscriptstyle GL}}^{-1}=8T_{\mathrm e}[x\ln(T_1/T_2)-\ln(T_{\mathrm c}/T_2)]/\pi$,
$T_{\mathrm c}$ is the equilibrium  superconducting transition temperature.

The net electric current has a standard form
$\mathbf{j}= e\nu \int_\epsilon [\mathbf{j}_{\mathrm e}(\epsilon)
-\mathbf{j}_{\mathrm h}(\epsilon)]/2$.
To derive the thermal current, we construct the gauge invariant vector out of
$\mathbf{j}$, $(\phi,\mathbf{A})$ and the energy current
$\mathbf{j}_E=\nu \int_\epsilon \epsilon[\mathbf{j}_e(\epsilon)+\mathbf{j}_h(\epsilon)]/2$:
 $\mathbf{j}_{\rm th}=\mathbf{j}_E-\phi \mathbf{j}$.
This agrees with the local equilibrium temperature gradient
form of $\mathbf{j}_{\rm th}$ and satisfies the conservation law,
$\bnabla \mathbf{j}_{\rm th}=\boldsymbol{\mathcal{ E}}\mathbf{j}$,
where $\boldsymbol{\mathcal{E}}$ is the electric field.
In a thin film, the gradient is reduced to the
difference at the interfaces, and
 in the setting of Fig.\ref{fig1}
\begin{gather}\label{eq:J_th}
j_{\mathrm {th}}=\frac1{2R_{\scriptsize \Sigma}e^2}
\left\{\frac{\pi^2}3(T_1^2-T_2^2)-e^2V_{\scriptscriptstyle G}^2(2x-1)\right\}.
\end{gather}
The same expression for the thermal current holds for the 2D film or
wire connected through the edges and/or ends.
Then the parameter $x$ is nothing but the coordinate along the layer (wire)
normalized by its width (length).

The behavior of subsystem temperatures as functions of the external drive
given by  Eqs.\,\eqref{eq:Tcp}-\eqref{eq:Te} is shown in Fig.\ref{fig1}.
Below, we concentrate on the case $R_1=R_2$ (i.e.~$x=1/2$),
since then the system is the most strongly driven out of equilibrium.
(Note that the equilibrium situation is realized for $x=0$ and $x=1$,
with temperature being $T_2$ and $T_1$, respectively.)
If transport is dominated by the thermal current due to the temperature
difference between the substrate and the gate,
and $\tau_{\mathrm{\scriptscriptstyle GL}}^{-1} \ll T_2\ll T_1$,
at zero bias $V_{\scriptscriptstyle G}=0$,
then at $x=1/2$ the quasiparticles and fluctuating Cooper pairs
are much colder than the electron-hole pairs:
$T_{\mathrm e} \approx 2T_2\backsimeq T_{\mathrm{\scriptscriptstyle Cp}}\ll
T_{\mathrm{eh}}\approx T_1/4$.
On the contrary, if the external drive is provided mostly by the voltage drop,
the electron-hole pairs appear to become the coldest electronic subsystem for $x=1/2$:
$T_{\mathrm e}\simeq T_{\mathrm{\scriptscriptstyle Cp}}\gg T_{\mathrm{eh}}$,
(see Fig.~\ref{fig1}b).
It reflects the fact that the electric current affects mostly charged particles,
rather than the neutral electron-hole pairs.
The relation $T_{\mathrm {\scriptscriptstyle Cp}}\simeq T_{\mathrm e}$ (see Fig.\ref{fig1}a)
holds at all $x$ as long as $V_{\scriptscriptstyle G}\lesssim\min(T_1,T_2)$ [since
 $z_{\mathrm{Cp}}$ is of order of unity in Eq.\,\eqref{eq:Tcp} at small voltages].
The single particle excitations may become much hotter than the Cooper pairs,
$T_{\mathrm e}\gg T_{\mathrm {\scriptscriptstyle Cp}}$, at elevated $V_{\scriptscriptstyle G}$
and away from $x=1/2$, where $z_{\mathrm{\scriptscriptstyle Cp}}$
significantly deviates from unity, (see Fig.\ref{fig1}b) \cite{footnote_Te}.
%%%%
%%%%

We derive the kinetic coefficients using the Keldysh functional integral technique.
The Keldysh partition function in the coherent state basis is defined as:
$\mathcal Z=\int D\overline{\psi}D\psi \exp\{iS[\overline{\psi},\psi]\}$, where
$S[\overline{\psi},\psi]=\int_{\cal C}dt\left\{\int_{\mathbf{r}}(\bar\psi_\alpha i
\partial_t{\psi_\alpha})-H[\overline{\psi},\psi]\right\}$.
Here $H=H_0+H_{\mathrm {int}}$ is the Hamiltonian,
${\cal C}$ is the Keldysh contour and $\alpha\equiv\uparrow$,$\downarrow$ is the spin variable.
The one-particle Hamiltonian
$H_0=-\int_{\mathbf r}\bar\psi_\alpha\left[{(\bnabla - ie\mathbf{A})^2}/{2m} +
\mu - U_{\mathrm {dis}}
-e\phi \right]\psi_\alpha$,
where $\mathbf A$, $\phi$ and $U_{\mathrm {dis}}$ are vector, scalar and disorder potentials,
and $\mu$ is the Fermi energy;
the tensor summation over the spin indices is implied.
The interaction Hamiltonian describes the electron-electron interaction in the singlet
and Cooper channels,
$H_{\mathrm {int}}=(1/2) \int_{\mathbf{r},\mathbf{r'}}\{\rho_{\mathbf{r}}
\Gamma^{(\rho)}_{\mathbf{r-r'}}\rho_{\mathbf{r'}}- (\lambda/\nu)\delta_{\mathbf{r-r'}}\bar
\psi_{\uparrow}\bar\psi_\downarrow\psi_\downarrow \psi_\uparrow\}$,
where $\Gamma^{(\rho)}$ is the singlet channel interaction amplitude,
$\rho(\mathbf r)=\bar\psi_\alpha(\mathbf r)\psi_\alpha(\mathbf r)$ is the local electron density.
Averaging over Gaussian disorder and carrying out the standard decoupling in the four-fermion
terms in the action via the Stratonovich-Hubbard fields $\{\Delta,Q,\Phi\}$~\cite{L_Kamenev},
 and integrating out the degrees of freedom with the energies higher than $1/\tau$,
 we arrive at the Keldysh nonlinear $\sigma$-model action:
\begin{gather}\label{eq:S}
S[\Qk,\Dk,\Fk]=S_{\Delta}+S_{\Phi}+S_{Q}.
\end{gather}
Here $S_{\Delta}=-\nu(1+\lambda)\Tr[\Dk^{\dagger}\check{Y}\Dk]/2\lambda$,
$S_{\phi}={e^2\nu}\Tr[\check b_{\mathcal{K}}\check{Y}\check b_{\mathcal{K}}]/2$,
	%\begin{align}\label{q:S_Q}
		$S_{Q}={i \pi \nu}\Tr[D (\partial_{\bf{r}}\Qk)^2/4- \Si\partial_t\Qk
		%\notag\\&
		+ie\check b_{\mathcal{K}}\Qk+i\Dk\Qk ]$ and $\check Q^2=1$.
%	\end{align}
We used the unitary limit $\Gamma^{(\rho)}_{\mathbf{r-r'}}\to\infty$
implying that the charge screening length in the electron liquid is much smaller than the mean free path.
The action \eqref{eq:S} holds while the fermionic and bosonic temperatures that
follow from it are much smaller than $1/\tau$.
The check mark above the field variable indicates that it is defined on the
tensor product of the Keldysh and Nambu spaces spanned by the Pauli matrices
$\hat{\sigma}_i$ and $\hat{\tau}_i$, $i\in\{0,x,y,z\}$, respectively.
So, $\check{Y}=\hat{\sigma}_x\otimes \hat{\tau}_0$, $\Si=\hat{\sigma}_0\otimes\hat{\tau}_z$
and $\check b=\check \Phi+\check\phi$.
Multiplication in time space is implicitly assumed,
and ``Tr'' includes an integration over real space.
The subscript $\cK$ denotes the gauge transformed fields:
$\Fk=\check{\Phi}-\partial_{t}\check{\cK}$,
and  ${\Ak}={\check{\mathbf{A}}} +{\bf{\bnabla}}{\check{\cK}}$,
where $\check{\cK}=[k^{cl}\hat{\sigma}_{0}+k^{q}\hat{\sigma}_{x}]\otimes\hat{\tau}_{0}$
[$\check A$ and  $\check \Phi$ are defined similarly].
Then $\check{\Delta}= [\Delta^{cl}\hat{\sigma}_{0}+
\Delta^{q}\hat{\sigma}_{x}]\otimes\hat{\tau}_{+}-\mathrm{h.c.}$,
and $\Dk(\br,t)=e^{ie \Si{\check{\cK}}(\br,t)}\check{\Delta}e^{-ie\Si{\check{\cK}}(\br,t)}$,
$\Qk$ is defined in the same way. The quantum (q) and classical (cl)
components are defined in the standard way as the half-sum and half-difference
of the field values at the lower and upper brunches of the Keldysh time-contour.
The field $\Delta$ becomes the superconducting order parameter
on the mean-field level,
while the saddle-point equation for $\check Q$ corresponds to the Usadel quasiclassical equations
where $\check Q$ plays the role of the quasiclassical Greens function.
The covariant spatial derivative is
$\partial_{\mathbf{r}}\check{Q}_{\cK}= \bnabla_{\mathbf{r}}\check{Q}_{\cK}-
ie[\check{\Xi}\check{\mathbf{A}}_{\cK},\check{Q}_{\cK}]$.

%%%%%%%%%%%%%%%%%%%%%%%%%
%%%%%%%%%%%%%%%%%%%%%%%%%
We use linear response formalism  to find the interaction corrections
to the kinetic coefficients of the electron liquid integrating
out the fluctuations around the metallic saddle point of $S$,
$\check \Lambda= \check{\mathcal{U}}(\hat{\sigma}_z\otimes\hat{\tau}_z) \check{\mathcal{U}}^{-1}$,
where
\begin{gather}\label{lambda}
\check{\mathcal{U}}_{t,t'}(\br)=\check{\mathcal{U}}_{t,t'}^{-1}(\br)=
\left(\begin{array}{cc}\delta_{t-t'-0}\hat{\tau}_0&
\hat{F}_{t,t'}(\br)\\
0&-\delta_{t-t'+0}\hat{\tau}_0\end{array}\right).
\end{gather}
Here $\hat{F}_{t,t'}(\br)=\diag(F_{t,t'}^e(\br),F_{t,t'}^h(\br))$.
Performing the Wigner transformation we map $F^{\mathrm{e(h)}}_{t,t'}$
to the quasiparticle distribution functions:
$F_\mathrm{e(h)}(\epsilon,\tau)\equiv 1-2f_{e(h)}(\epsilon,\tau)$.
We split the gauge field, $k$, and electromagnetic fields
 into the slow, $\langle k\rangle$, and fast, $\delta k$, components.
The fast components describe the fluctuations in the electron system
and slow components are related to the gauge transformations
of the external field potentials~\cite{Chtch_PRL_2008}.
To optimize the fluctuations of the quasiparticle phases induced by
the Coulomb forces  \cite{L_Kamenev,Chtch_PRL_2008}
we solve the equation that couples $\delta k$ with the fluctuating electromagnetic fields:
$\delta b^{\mathrm q}_{\cK}-D\bnabla\, \delta\mathbf{g}^{\rm q}=0$,
\begin{gather}%\label{eq:cond_2_Fourier1}
D\bnabla\delta\mathbf{g}^{\mathrm{cl}}(\br,\omega)+\delta b^{\mathrm{cl}}_{\cK}(\br,\omega)
=-2 (1+2N_\omega) D\bnabla\delta
\mathbf{g}^q(\br,\omega)\,,
\end{gather}
where $\delta b^{\mathrm{q}(\mathrm{cl})}_{\mathcal{K}}=\delta\Phi^{\mathrm{q}(\mathrm{cl})}-
\partial_t \delta k^{\mathrm{q}(\mathrm{cl})}$, $\delta \mathbf{g}^{\mathrm{q}(\mathrm{cl})}
=\bnabla\delta k^{\mathrm{q}(\mathrm{cl})}$.
This way we fix the gauge and define $N_{\omega}$ as given by Eq.~(\ref{eq:1223}).
In the local equilibrium,
$f^{e(h)}(\epsilon,\tau)=f_F(\epsilon\mp e(\phi-\partial_\tau\langle k^{\mathrm{cl}}\rangle))$,
and then choosing $\langle k^{\mathrm{cl}}\rangle=\phi t$
one obtains the gauge invariant distribution functions.
%%%%%%%%%%%%%%%%%%%%%%%%%%%%%%%%
%%%%%%%%%%%%%%%%%%%%%%%%%%%%%%%%

Now, the electric conductivity is obtained by differentiating the partition function over
the quantum and classical components of the vector potential, and then integrating out the gauge, $\Delta$- and $Q$-fluctuations (diffusion and Cooperon
degrees of freedom). The major fluctuation contribution to
the in-plane conductivity close to the transition is the Maki-Thompson correction:
\begin{multline}\label{eq:dsMT}
\delta\sigma^{\mathrm{MT}}\approx \frac{iDe^2}{2\pi}\int_{\bf{q},\omega,\epsilon} [L^{-1}]_K\times
\\
\frac{|L_{R,\omega}(q)|^2 \partial_{\epsilon}f_{\mathrm e}(\epsilon)+|L_{R,-\omega}(q)|^2\partial_{\epsilon} f_{\mathrm h}(\epsilon)}{D^2q^4+(2\epsilon-\omega)^2},
\end{multline}
where $[L^{-1}]_K\approx i\pi T_{\mathrm {\scriptscriptstyle Cp}}/
(2 T_{\mathrm {e}} z_{\mathrm {\scriptscriptstyle Cp}})$,
$L^{-1}_{R,\omega}=\pi[-(\tau_{\mathrm{GL}}z_{\mathrm{Cp}})^{-1}
+(-Dq^2+i\omega-2i e \phi_{\mathrm{f}}-i 4 T_e (1-2 \tilde{f}_{\scriptscriptstyle{R}}(0)))
(1+i T_{\mathrm{e}}\Omega^{-1})]/(8T_{\mathrm{e}})$.
In two dimensions the integral \eqref{eq:dsMT} diverges logarithmically  at small $q$,
and should be cut off at $1/\tau_\phi$ leading to the result \eqref{eq:2},
for $\tau_{\mathrm {\scriptscriptstyle GL}}\ll \tau_{\phi}=\sigma_{\scriptstyle{D}}d\pi^2\left(\epsilon_0^2+\tau_{\mathrm {\scriptscriptstyle GL}}^{-2}\right)/\left(8e^2T_{\mathrm {\scriptscriptstyle Cp}}T_{e}f(\epsilon_0)\right)$. Here $f(\epsilon_0)=\left(\pi|\epsilon_0|-2\epsilon_0\arctan[\tau_{\mathrm {\scriptscriptstyle GL}}^{-1}\epsilon_0^{-1}]+\tau_{\mathrm {\scriptscriptstyle GL}}^{-1}\log[4\tau_{\mathrm {\scriptscriptstyle GL}}^{-2}/\left(\epsilon_0^2+\tau_{\mathrm {\scriptscriptstyle GL}}^{-2}\right)]\right)$, where $\epsilon_0=4T_e(1-2\tilde{f}_{\scriptscriptstyle R}(0))-\tau_{\mathrm {\scriptscriptstyle GL}}^{-1}T_e/\Omega.$
Corrections from the fluctuating Cooper pairs are suppressed
by the magnetic field perpendicular to the film, and the Aronov-Altshuler
correction becomes the dominating one:
\begin{gather}
\delta\sigma^{\mathrm{\scriptscriptstyle AA}}\approx \frac{2e^2D}{\pi}\Imag\int_{\mathbf{q},\omega}
\frac{\partial_{\omega}\left[\omega(1+2N_\omega)\right]}{(Dq^2+i\omega)^2}.
\end{gather}
This integral also diverges logarithmically in 2D, and the
integration over $\omega$ should be cut off at the upper limit
by $1/\tau$ \cite{Aronov-Altshuler},
 whereas the infrared boundary is determined by $T_{\mathrm{eh}}$ \cite{footnote},
giving rise to Eq.~\eqref{eq:1}\cite{footnote_RG}.

Next, we calculate the thermal Nyquist noise. In the noninteracting case, the current density  $\mathbf{j}=\sigma_{\mathrm{\scriptscriptstyle D}}[\boldsymbol \zeta +\mathcal E]$. Here $\boldsymbol \zeta$ is the random electrical field induced
by the fluctuations of the vector potential quantum components,
the statistical properties of which are defined by the Fourier transform
$\langle\zeta^2\rangle_\omega=\omega (1+2N_\omega)/\sigma_{\mathrm{\scriptscriptstyle D}}$.
Then, we obtain the Nyquist noise $S_{\mathrm{th}}(\omega)= \sigma_{\mathrm{\scriptscriptstyle D}}\omega (1+2N_\omega)$.
This result is derived in the limit $V\ll \omega$.
If $\omega\ll T_{\rm eh}$ then $[1+2N_\omega]\to 2T_{\rm eh}/\omega$,
and we get Eq.~(\ref{eq.3}).

Fermionic and bosonic  temperatures can be straightforwardly detected,
by direct measurements of the magnetoresistance and temperature
dependence of the resistance at fixed
magnetic fields  using the setup
shown in Fig.\,1.
Varying $T_1$ and $T_2$ and the gate voltage, one can directly observe the effect of
$T_{\mathrm{\scriptscriptstyle Cp}}$ in the Maki-Thompson correction.
At the same time changing temperatures $T_1$ and $T_2$ at the fixed high magnetic field,
where the Maki-Thompson contribution is suppressed,
one infers the information about $T_{\mathrm{eh}}$.
Also $T_{\mathrm eh}$  can be determined
from the Nyquist noise measurements.

The work was funded by the U.S. Department of Energy Office of Science
through the contract DE-AC02-06CH11357, by RFBR, Dynasty and by the RF President foundation.


\begin{thebibliography}{99}
\bibitem{footnote_Te}
To measure $T_{\mathrm{e}}$ one should filter out the current carried by low-energy electrons,
e.g. by attaching to the film a quantum dot with the resonance level tuned to $\epsilon=0$.
\bibitem{footnote}
We find that $N_\omega$ is very well approximated
by the Bose-function at $T=T_{\mathrm {eh}}$.
\bibitem{footnote_RG}
The derivation of Eq.\eqref{eq:1}
is a step towards the generalization of the equilibrium Finkelstein renormalization
group\cite{FinkelsteinReview} (RG) onto the far from equilibrium case.
One can show~\cite{forthcoming} that being far from the equilibrium
one has to replace the equilibrium RG flow
parameter $\Lambda(T)=\sqrt{D/z T l^2}$ by
%the nonequilibrium flow quantity
$\Lambda(T_{\rm eh})$, where $l$ is the mean free path and $z$ is the interaction
energy (effective mass) renormalization factor equal to unity at high temperatures.
\end{thebibliography}
\end{document}